\documentclass[10pt,conference]{IEEEtran}

\usepackage{graphicx}
\usepackage{url}
\usepackage{listings}
\usepackage{pxfonts}
\usepackage{xcolor}
\usepackage{booktabs}
\usepackage{hyperref}

\begin{document}

\title{GE852: A Dataset of 852 Game Engines}

\author{
\IEEEauthorblockN{Chaitanya S. Lakkundi}
\IEEEauthorblockA{\textit{Indian Institute of Technology} \\
Tirupati, India \\
cs18s502@iittp.ac.in}
\and
\IEEEauthorblockN{Vartika Agrahari}
\IEEEauthorblockA{\textit{Indian Institute of Technology} \\
Tirupati, India \\
cs18m016@iittp.ac.in}
\and
\IEEEauthorblockN{Sridhar Chimalakonda}
\IEEEauthorblockA{\textit{Indian Institute of Technology} \\
Tirupati, India \\
ch@iittp.ac.in}
}
\maketitle

\begin{abstract}

Game engines provide a platform for developers to build games with an interface tailored to handle the complexity during game development. To reduce effort and improve quality of game development, there is a strong need to understand and analyze the quality of game engines and their various aspects such as API usability, code quality, code reuse and so on. To the best our knowledge, we are not aware of any dataset that caters to game engines in the literature. To this end, we present GE852, a dataset of 852 game engine repositories mined from GitHub in two languages, namely Java and C++. The dataset contains metadata of all the mined repositories including commits, pull requests, issues and so on. We believe that our dataset can lay foundation for empirical investigation in the area of game engines.

\end{abstract}


\maketitle

\section{Introduction} \label{introduction}



Games are considered as one of the sophisticated and complex forms of software \cite{washburn2016went}\cite{blow2004game}. Studies indicating positive effects of playing video games has opened up new opportunities for game developers to develop innovative games \cite{granic2014benefits}. This has also led to a drastic increase in the number of games available in platforms such as \textit{Google PlayStore}\footnote{\url{https://play.google.com/store?hl=en}} and \textit{Steam}\footnote{\url{https://store.steampowered.com}}. The rise in the number of games led to diversified ways of game development, and researchers have noted that the process of game development is different from software development \cite{murphy2014cowboys, santos2018computer}. On the other hand, game software development is considered as an effort-intensive activity \cite{aleem2016game}. A recent study by Kasurin et al. \cite{kasurinen2017concerns} reveals several concerns of developers in game development such as code reusability, API dependencies, compatibility, maintenance and so on.



 Game engines provide developers with reusable software development kits with inbuilt APIs and functionality which helps developers perform various tasks more efficiently \cite{hudlicka2009affective}.  Paul et al. \cite{paul2012history} discussed different types of game engines, their specifications and features in detail. The development of game engines itself is considered as a complex software development process \cite{lewis2002game} with complex implementation \cite{gregory2014game}. However, the increasing number of game engines makes it difficult for game developers to choose an appropriate game engine that is suitable in the context of their game development \cite{paul2012history}. It is also critical for game developers to analyze game engines for appropriate API compatibility, software evolution, dependencies, code clones and reuse and several issues that align their own software development process. Existing empirical studies on game engines have largely focused on performance evaluation. Messaoudi et al. \cite{messaoudi2015dissecting, messaoudi2017performance} have presented a study on performance evaluation of the CPU and GPU of different modules in Unity3D. However, there is a strong need to perform empirical research on different aspects of game engines such as energy efficiency, code clones and reuse, API usage patterns, release engineering and pull requests among others to support game developers.

While there exist several datasets to support empirical research in the broad areas of software engineering such as API usage patterns \cite{zhong2009mapo, wang2013mining}, energy consumption \cite{capra2012software, linares2014mining}, code clones \cite{roy2009comparison}, source code metrics \cite{haefliger2008code} and so on, to the best of our knowledge, there have been no datasets that provide game engines for empirical research. With this background, we present \textbf{GE852}, a dataset\footnote{The dataset is available online at \url{http://bit.ly/GameEngines852}} of game engines consisting of 852 distinct game engines mined from GitHub in two programming language, Java and C++. Total number of game engines written in Java are 408 and that of C++ are 444. Overall, \textbf{GE852} contains metadata of 2627 game engines including the forked projects. We believe that our dataset will help in fostering new direction in domain of game engines.

The remainder of this paper is structured as follows. In Section \ref{data_collection}, we discuss the data collection process and Section \ref{database_schema} details database schema. Section \ref{dataset_usage} lists the potential usages of the dataset for researchers and practitioners. Finally, we discuss limitations in Section \ref{limitations} followed by conclusions and future work in Section \ref{conclusion}.


 \begin{figure*}
     \centering 
     \includegraphics[height=5cm,keepaspectratio]{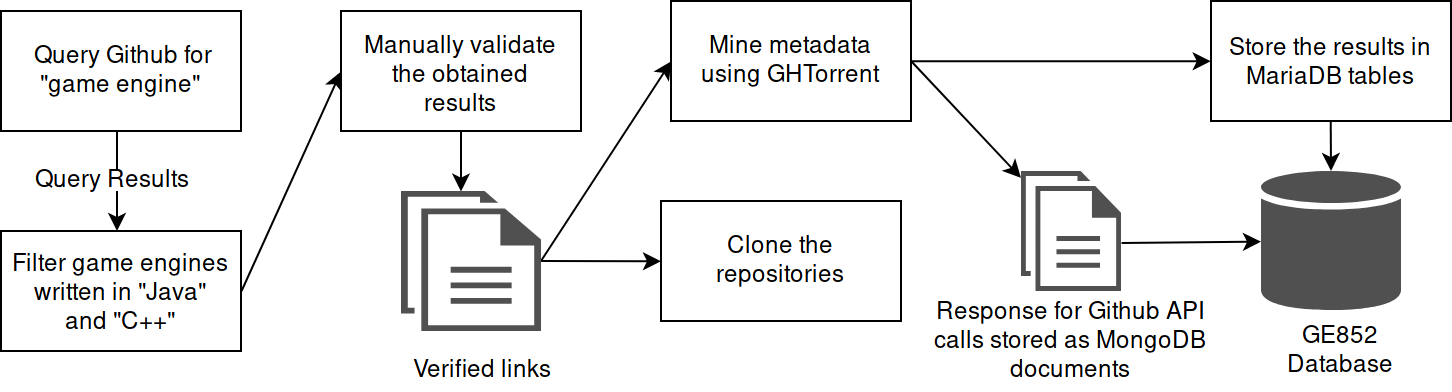}
     \caption{Flow of extraction process}
      \label{fig:pattern_pipeline}
 \end{figure*}


\section{Data Collection} \label{data_collection}

The data extraction process is shown in Figure \ref{fig:pattern_pipeline} and is elaborated here. The process of collection of metadata began with preliminary automated filtering, followed by manual validation and ended with mining the actual repositories using GHTorrent\footnote{\url{http://ghtorrent.org}}, that is widely used for mining GitHub repositories \cite{Gousi13}.

We enumerate the following steps followed to create the dataset.

\textbf{Step 1 - Collection of game engine repository links.}
As an initial step, we queried GitHub with keywords, ``game engine'' that returned over 29 thousand repositories.

\textbf{Step 2 - Filtration of game engines by programming language.}
We then filtered out game engines written in Java and C++.


\textbf{Step 3 - Manual validation of the obtained results.}
We noticed that even after filtration, some repositories did not truly belong to the game engine category that we were intending to download. An instance of this was initially noticed in the second search page of Java game engines when sorted by most stars. We found that the repository ``arjanfrans/mario-game'' was listed along with the surrounding game engine repositories. Then, we anticipated many more false positives and manually validated every listed repository by reading the description of every repository. Then we had a corpus of GitHub links pointing to game engine repositories.

\textbf{Step 4 - Collection of metadata.}
The collection of metadata from GitHub could be done either using custom made scripts or using any open source tool such as PyDriller \cite{spadini2018pydriller}, RepoDriller\footnote{\url{https://github.com/mauricioaniche/repodriller}}, GitMiner\footnote{\url{https://github.com/UnkL4b/GitMiner}} and Boa \cite{dyer2013boa}. We found that GHTorrent is a tool that has been used by researchers to mine data from GitHub since 2012 and continuously lists the daily dumps. For our study we independently mined data using GHTorrent without using the dumps provided by them. 




\textbf{Step 5 - Store GitHub API responses in MongoDB.}

Finally, we stored the raw data as MongoDB documents and structured data in MariaDB (a fully open source fork of MySQL) tables. The raw data contains responses to every REST API call. 

\textbf{Step 6 - Dropping irrelevant tables.}
GHTorrent, by default retrieves the metadata with fields as described by their schema\footnote{\url{http://ghtorrent.org/relational.html}}. Tables and attributes relevant to the context of game engines were retained and rest of them were dropped. We considered those tables that are relevant for conducting empirical research in the context of game engines. 


\begin{figure*}[ht]
   \centering
   \includegraphics[width=6in, height=4in]{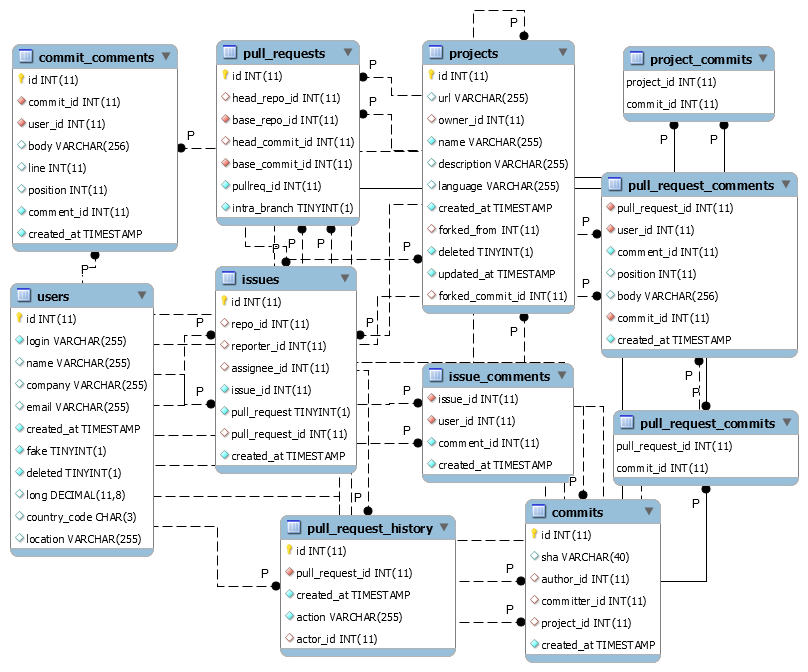}
    \caption{Database Schema}
    \label{fig:schema}
\end{figure*}


\section{Database Schema} \label{database_schema}
The schema of the database is shown in Figure \ref{fig:schema}. The description of all the 11 tables are given below:
\begin{itemize}
    \item \textbf{projects}: This table contains attributes such as {\fontfamily{qcr}\selectfont
name, url, descriptor, language, owner\_id} that describe preliminary information about the repository. 
The {\fontfamily{qcr}\selectfont
forked\_from} field indicates the project\_id of the project from where it has been forked. It contains a NULL value for repositories that have not been forked. The {\fontfamily{qcr}\selectfont
deleted} field indicates if the repository has been deleted from GitHub.

\item \textbf{users}: It gives the information of all the users related to game engine projects. It contains various attributes depicting the information about the users such as {\fontfamily{qcr}\selectfont
name, email, login, company, country\_code}. The {\fontfamily{qcr}\selectfont
fake} attribute states whether the user is real or fake. Real users can own projects, push commits and create pull requests whereas fake users can only appear as authors or committers of commits. There is another attribute with name {\fontfamily{qcr}\selectfont
deleted}, which indicates an earlier presence of the user on GitHub whose details no longer exist.

\item \textbf{commits}: {\fontfamily{qcr}\selectfont
project\_id} refers to the project to which a particular commit has been associated first, which might not be the project\_id of the project it was initially pushed to. Another attribute of this table is {\fontfamily{qcr}\selectfont
sha}, which acts as global identifier for each commit. 

\item \textbf{projects\_commits}: It stores the relation between commits and projects, as more than one project can share the same commit if one is a fork of another.

\item \textbf{commit\_comments}: It stores the comments associated with each commit. If a commit has association with any pull request, then its comment can be obtained from the {\fontfamily{qcr}\selectfont
pull\_Request\_Comments} table.

\item \textbf{issues}: Field {\fontfamily{qcr}\selectfont
id} is the primary key, whereas, {\fontfamily{qcr}\selectfont
repo\_id} is the foreign key pointing to the {\fontfamily{qcr}\selectfont
projects} table.
{\fontfamily{qcr}\selectfont
assignee\_id} refers to the user to whom the issue was assigned at the time of creating the issue. {\fontfamily{qcr}\selectfont
issue\_id} is unique identifier given by GitHub to each issue.
Fields {\fontfamily{qcr}\selectfont
pull\_request} and {\fontfamily{qcr}\selectfont
pull\_request\_id} are for pull\_requests associated to any issue. 
Creation date of the repository is stored in {\fontfamily{qcr}\selectfont created\_at} column.

\item \textbf{issue\_comment}: It stores information about discussions being made in the issues section in GitHub. Each comment is related to a issue by {\fontfamily{qcr}\selectfont
issue\_id}. {\fontfamily{qcr}\selectfont
user\_id} uniquely identifies the user who made the comment.

\item \textbf{pull\_requests}:
A pull request is initiated by ({\fontfamily{qcr}\selectfont
head\_repo\_id}, {\fontfamily{qcr}\selectfont
head\_commit\_id}) to ({\fontfamily{qcr}\selectfont
base\_repo\_id}:{\fontfamily{qcr}\selectfont
base\_commit\_id}). Thus, head and base information is stored in these attributes. {\fontfamily{qcr}\selectfont
pull\_request\_id} is a unique GitHub identifier to pull requests. {\fontfamily{qcr}\selectfont intra\_branch} indicates whether the head and base repository of pull request are same or different.

\item \textbf{pull\_request\_comments}: It stores all the comments being made on commits associated with any pull request.

\item \textbf{pull\_request\_commits}: stores the association of a commit with a pull request.
\item \textbf{pull\_request\_history}: This table contains the history of  pull\_requests.
\end{itemize} In Figure \ref{fig:schema}, the dotted and solid lines depict the connections between different tables in the database schema. The solid line denotes an identifying relationship which means that primary key of parent entity is included in primary key of child entity. However, the dotted line refers to a non-identifying relationship which indicates that primary key of the parent table is included in child entity but not as part of its primary key.

\section{Dataset Usage or Applications} \label{dataset_usage}
\begin{table*}[ht!]
\centering
\caption{API usage patterns found in Open Diablo Game Engine using PAM tool}
\begin{tabular}{l|l}
\multicolumn{1}{c|}{\textbf{API Patterns}}                                                                              & \multicolumn{1}{c}{\textbf{Probability of usage}} \\
\midrule
SDL2.SDL.UTF8\_ToNative & 0.21983  \\
SDL2.SDL.UTF8\_ToManaged, SDL2.SDL.SDL\_free                                                        & 0.15948  \\
OpenDiablo2.Common.Models.Mobs.Stat.GetCurrent, OpenDiablo2.Common.Models.Mobs.Stat.GetMax
        & 0.10345  \\
OpenDiablo2.Common.Interfaces.Mobs.IStatModifier.GetValue & 0.10345   \\
OpenDiablo2.Common.Interfaces.IRenderWindow.LoadSprite      & 0.03017             
\end{tabular}
    \label{tab:example_opendiablo_engine_patterns}
\end{table*}
The main goal of GE852 dataset is to create a platform for researchers and practitioners to empirically investigate research challenges in the broad areas of software engineering, that were largely undermined in the context of game engines. We enumerate a set of preliminary research directions based on our dataset.

\subsection{Discovering API usage patterns of game engines}

There has been existing research on API usability for software development \cite{zibran2011useful}. However, it is important to understand API usage in game engines, which is not studied till now. Through our dataset, we have conducted a preliminary experiment to extract API usage patterns and the probability of the occurrence of APIs along with other APIs. Table 1 shows some statistics for a game engine that were calculated using a tool called PAM (Probabilistic API-miner) \cite{fowkes2016parameter}. The table shows that the API \textit{OpenDiablo2.Common.Models.Mobs.Stat.GetCurrent} is used more frequently than the API \textit{OpenDiablo2.Common.Interfaces.IRenderWindow.LoadSprite}. We see that there is scope for doing extensive research in this direction and as a first step we are working on determining API use and misuse in game engines. 

\subsection{Analyzing pull-requests and commits}
Rahman et al. \cite{rahman2014insight} have emphasized the need to study successful and unsuccessful pull-requests. The GE852 dataset could be potentially used for understanding potential reasons for acceptance or rejection of pull-requests and its relationship to developer and project-specific information. Further to this, we can study pull requests across game engine repositories and also issues. We can analyze the commit history of the project to infer insights that could help developers. We can also analyze code changes across the version history of the game engines. 



\subsection{Issues in game engines}

Bissyande et al. \cite{bissyande2013got} emphasize that issues play a critical role in improving a project's performance. Issues help software developers to understand flaws in specific modules of software development, their source, potential reasons behind issues and eventually what measures could to be taken to resolve these issues. There is a need to empirically investigate issues in the context of game engines, as they form the basis for further development of large number of games. Hence, we see that our dataset could be used to conduct empirical studies in understanding, analyzing and reporting issues in game engines. 


\subsection{Energy Efficiency}

In the last few years, there is an ever increasing demand for energy-efficient apps and games, more so for mobile platforms. This presents a tremendous opportunity for conducting empirical research that analyzes game engines for energy efficiency, which is largely ignored in the current literature. 


\subsection{Code quality and cloning}

Code quality and code clones are an active area of research in games \cite{roy2009comparison,cordy2011nicad}. However, we are not aware of existing research on analyzing code clones, code reuse and code quality patterns in game engines, which could be conducted through our dataset. 




\section{Limitations}
\label{limitations}
%
\begin{itemize}
    \item In the current dataset, we consider game engines written in Java and C++ only. We intend to extend the dataset to include game engines in other languages such as C\# and Javascript.
    \item Even after careful examination, some unintentional noisy data may have been provided in the dataset. We plan to refine the process and extend the dataset in our future works.
\end{itemize}



\section{Conclusion and Future Work} \label{conclusion}

We have presented GE852 as a dataset consisting of 852 game engines mined from GitHub. It essentially contains metadata such as commits, pull requests and issues of the respective game engines. To the best of our knowledge, this is the first dataset made available in the domain of game engines. We believe that this dataset will help in conducting a number of empirical studies on game engines. The quality of game engines may be analyzed that will lead to better quality of game engines. This dataset will help in exploring the underlying causes of why developers choose a particular game engine to develop their game over others. Moreover, this data can help in building tools which can benchmark game engines in terms of API usage, structural usability of game engines and performance. Thus, we believe that GE852 will open new opportunities and research work in the domain of game engines.

\bibliographystyle{IEEEannot}
\bibliography{gameengine.bib}

\end{document}